\documentclass[aps]{revtex4}
\usepackage{amsmath}
\usepackage{amssymb}
\newcommand\Ac{{\cal A}}
\newcommand\Bc{{\cal B}}
\newcommand\half{\scriptstyle{\frac{1}{2}}}
\newcommand{\proj}[2]{\left\vert{#1}\right\rangle\left\langle{#2}\right\vert}

\begin{document}
\title{Comment on `Underlining some limitations of the statistical formalism
       in quantum mechanics' by Fratini and Hayrapetyan}    
\author{Andr\'as Bodor$^1$}
\affiliation{
Department of Physics of Complex Systems, E\"otv\"os University\\
H-1117 Budapest, P\'azm\'any P\'eter s\'et\'any 1/A, Hungary}
\author{Lajos Di\'osi$^2$}
\email{diosi.lajos@wigner.mta.hu}
\homepage{www.rmki.kfki.hu/~diosi} 
\affiliation{
Wigner Research Center for Physics\\
H-1525 Budapest 114, POB 49, Hungary}
\date{\today}

\begin{abstract}
We point out that Fratini and Hayrapetyan ignored the randomness of mixing, 
a basic request to prepare a statistical ensemble rather than an ensemble in general. 
Their analysis is irrelevant for standard statistical ensembles, their conclusions 
about the limitations of standard theory become unjustified.
\end{abstract}


\maketitle
Fratini and Hayrapetyan (FH) claimed recently that ``ensembles of states 
which are represented by the same density operator in quantum mechanics
can behave differently in experiments'' \cite{FraHay11}. To prove this,
they choose two totally unpolarized ensembles $\Ac$ and $\Bc$ of spin-1/2 
particles. $\Ac$ is made of an equal number $N/2$ of particles 
having full polarization $\hbar/2$ and $-\hbar/2$, respectively, along the 
$x$ axis. $\Bc$ is 
prepared the same way just using particles fully polarized along the $z$ axis.
FH claim correctly that $\Ac$ and $\Bc$ behave differently in
experiments. But their conclusion that ``the statistical description of an ensemble
of states, as given by its density matrix, ..., should be considered as incomplete''    
is, to our opinion, incorrect.
 
FH are apparently unaware of their ensembles are different from what 
standard statistics as well as standard quantum mechanics understand as {\it statistical ensembles}.
FH prepare the ensemble $\Ac$ of an \emph{equal number} $N/2$ of 
$\pm\hbar/2$ polarizations which is obviously no legitimate construction for 
a {\it statistical ensemble}.
Statistical ensembles must consist of {\it independent} states, this requires 
\emph{random} mixing \cite{ranmix}. 
Statistical formalism will then attribute a {\it unique} statistical ensemble to the
unpolarized state, different from both ensembles $\Ac$ and $\Bc$ of FH.  

Totally unpolarized statistical ensembles are extensively used in
quantum information theory. If the statistical (i.e.: random) mixture of 
$\vert S_z,\pm1\rangle$ were distinguishable from the statistical mixture of $\vert S_x,\pm1\rangle$ 
then standard information and communication protocols \cite{NieChu00,Dio11} 
would not work at all --- they are just based on the fact that these two statistical
ensembles are identical. 

The mistake of FH has nothing to do with quantum mechanics rather it is a 
classical statistical misconception. If we mix $N/2$ black balls and
$N/2$ white balls in an urn then the emerging ensemble is not a statistical
ensemble. To prepare the statistical ensemble, we should have mixed N black 
or white balls \emph{chosen at random} each. The resulting statistical ensemble 
will contain black and white balls according to the balanced binomial distribution:
$$
p(N_{black})=\left({N\atop N_{black}}\right)2^{-N}.
$$
This statistical ensemble is what standard statistics assigns 
to the uniform distribution of black and white balls. The urn containing
an equal number of them is not a statistical ensemble in the above sense, it does not
represent the uniform probability distribution. 
To be sure, the distribution of black balls is peaked at $N/2$:
$$
p(N_{black})=\delta_{N_{black},N/2}
$$
and this is sharply different from the previous correct one.
Therefore, in the spirit of \cite{FraHay11}, we could have proposed 
similar (mistaken) limitations of classical statistical formalism: 
Since two ensembles with the same one-ball probability distribution 
behave differently in experiments, the ensembles can not be fully 
characterized by their one-ball distributions. We would obviously
be wrong: one of the ensembles in question is not a legitimate 
statistical ensemble, its experimental difference from the correct 
statistical ensemble is neither a surprise nor a point to claim 
limitations of standard statistics.

Quite similarly, neither $\Ac$ nor $\Bc$ are correct
statistical ensembles, their experimental difference from each other (and from
the correct statistical ensemble!) is not a surprise, the ensembles $\Ac,\Bc$
don't provide any limitation or alteration to the standard statistical 
formalism nor to the role of the density operator. 
Had FH used the correct mixing to construct the two ensembles they would have
left with no measurable difference between them which fact is fundamental in
the quantum theory and is particularly well understood in quantum
informatics.

To characterize the similarity and difference of the two ensembles $\Ac$ and $\Bc$,
let us note, that the one-particle density matrix of both ensembles is
identical to $\varrho_1={\half}I$, i.e., $1/2$ times the unit matrix: $\varrho^{\Ac}_1=\varrho^{\Bc}_1={\half}I$.
It represents the completely unpolarized state, indeed, as it should.
The two-particle density matrix of the correct statistical ensemble should be \cite{multirho}
$$
\varrho_2=\frac{1}{4} I\otimes I.
$$
The two-particle density matrices of the ensembles of FH differ from each other:
\begin{equation*}
\varrho^{\Ac}_2 = \frac{\half N(\half N-1)}{N(N-1)}\left(\proj{S_x^{++}}{S_x^{++}}+\proj{S_x^{--}}{S_x^{--}}\right)+ 
 \frac{N^2/4}{N(N-1)}\left(\proj{S_x^{+-}}{S_x^{+-}}+\proj{S_x^{-+}}{S_x^{-+}}\right) 
\end{equation*}
\begin{equation*}
\begin{split}
\varrho^{\Bc}_2 =& \frac{\half N(\half N-1)}{N(N-1)}\left(\proj{S_z^{++}}{S_z^{++}}+\proj{S_z^{--}}{S_z^{--}}\right)+ 
 \frac{N^2/4}{N(N-1)}\left(\proj{S_z^{+-}}{S_z^{+-}}+\proj{S_z^{-+}}{S_z^{-+}}\right) \\
 =& \frac{\half N(\half N-1)+N^2/4}{2N(N-1)} \left(\proj{S_x^{++}}{S_x^{++}}+\proj{S_x^{+-}}{S_x^{+-}}+ 
  \proj{S_x^{-+}}{S_x^{-+}}+\proj{S_x^{--}}{S_x^{--}}\right) +\\
 & \frac{\half N(\half N-1)-N^2/4}{2N(N-1)} \left(\proj{S_x^{++}}{S_x^{--}}+\proj{S_x^{--}}{S_x^{++}}+ 
 \proj{S_x^{+-}}{S_x^{-+}}+\proj{S_x^{-+}}{S_x^{+-}}\right)
\end{split}
\end{equation*}
where $\left\vert{S_x^{++}}\right\rangle=\left\vert{S_x,+1}\right\rangle\otimes\left\vert{S_x,+1}\right\rangle$, 
$\left\vert{S_x^{+-}}\right\rangle=\left\vert{S_x,+1}\right\rangle\otimes\left\vert{S_x,-1}\right\rangle$, e.t.c., 
stand for the corresponding two-particle states. The coefficients $\frac{\half N(\half N-1)}{N(N-1)}$ and
$\frac{N^2/4}{N(N-1)}$ are $1/2$ times the relative frequency of particle pairs with respectively
parallel and anti-parallel polarization.
In the second equation we used the identites like, e.g.:
\begin{equation*}
\proj{S_z^{++}}{S_z^{++}} = 
\frac{1}{2}\left(\proj{S_x^{++}}{S_x^{++}} + \proj{S_x^{--}}{S_x^{--}}\right).
\end{equation*}
These density matrices also differ from the form $\varrho_2=\frac{1}{4} I\otimes I$ of the correct statistical
ensemble. While the one-particle density matrices in $\Ac$,$\Bc$ and in the correct statistical ensemble 
are identical, $\varrho^{\Ac}_1=\varrho^{\Bc}_1=\varrho_1$, this is not true for the two- and multi-particle
density matrices. The variances, considered by FH, are intrinsically multi-particle
quantities, so the N-particle density matrix must be used in their computation. Therefore the outcomes for $\Ac$,
for $\Bc$ and for the standard statistical ensemble will differ from each other exactly as standard 
quantum mechanics predicts.

\vskip 25pt
The authors thank Tam\'as Geszti for his valuable remarks.
This work was supported by the Hungarian OTKA Grant No. 75129.


\begin{thebibliography}{99}
\bibitem{multirho} If $\rho=\rho_1$ is the (one-particle) density matrix, the multi-particle density matrices
of the corresponding statistical ensemble are $\rho_2=\rho\otimes\rho$, $\rho_3=\rho\otimes\rho\otimes\rho$, e.t.c..
\end{thebibliography}
\end{document}